\documentclass[aps,pre,twocolumn,
superscriptaddress,
showpacs,
 amsmath,amssymb,
]{revtex4-1}

\usepackage{graphicx}
\usepackage{dcolumn}
\usepackage{bm}

\usepackage{url}
\usepackage{color}
\numberwithin{equation}{section} 

\usepackage{setspace}
\usepackage{natbib}

\usepackage{lipsum} 
\usepackage[version=4]{mhchem}
\usepackage{siunitx}
\usepackage{ulem}
\DeclareSIUnit\Molar{M}
\DeclareSIUnit\Molar{\textsc{m}}
\DeclareSIUnit\cells{cells}
\DeclareSIUnit\gram{g}
\DeclareSIUnit\rpm{rpm}
\DeclareSIUnit\rcf{rcf}
\DeclareSIUnit\week{week}
\DeclareSIUnit\molecule{molecule}
\DeclareSIUnit\einstein{E}

\DeclareMathOperator{\Lapl}{\mathcal{L}}

\newcommand{\BXII}{B${}_{\text{12}}$}

\begin{document}

\preprint{APS/123-QED}

\title{Microbial mutualism at a distance: the role of geometry in diffusive exchanges}

\author{Fran\c cois J. Peaudecerf}
\email{peaudecerf@ifu.baug.ethz.ch}
\altaffiliation{Current address: Institut f\"ur Umweltingenieurwissenschaften, ETH Z\"urich, Stefano-Franscini-Platz 5, 8093 Z\"urich, Switzerland}
\affiliation{Department of Applied Mathematics and Theoretical Physics, University of Cambridge, Wilberforce Road, Cambridge CB3 0WA, United Kingdom}
\author{Freddy Bunbury}
\affiliation{Department of Plant Sciences, University of Cambridge, Downing Street, Cambridge  CB2 3EA, United Kingdom}
\author{Vaibhav Bhardwaj}
\affiliation{Department of Plant Sciences, University of Cambridge, Downing Street, Cambridge  CB2 3EA, United Kingdom}
\author{Martin A. Bees}
\affiliation{Department of Mathematics, University of York, Heslington, York  Y010 5DD, United Kingdom}
\author{Alison G. Smith}
\affiliation{Department of Plant Sciences, University of Cambridge, Downing Street, Cambridge  CB2 3EA, United Kingdom}
\author{Raymond E. Goldstein}
\email{R.E.Goldstein@damtp.cam.ac.uk}
\affiliation{Department of Applied Mathematics and Theoretical Physics, University of Cambridge, Wilberforce Road, Cambridge CB3 0WA, United Kingdom}
\author{Ottavio A. Croze}
\email{oac24@cam.ac.uk}
\affiliation{Cavendish Laboratory, University of Cambridge, J. J. Thomson Avenue, Cambridge CB3 0HE, United Kingdom}

\date{\today}

\begin{abstract}
The exchange of diffusive metabolites is known to control the spatial patterns formed by microbial populations, as revealed by recent studies in the laboratory. However, the matrices used, such as agarose pads, lack the structured geometry of many natural microbial habitats, including in the soil or on the surfaces of plants or animals. Here we address the important question of how such geometry may control diffusive exchanges and microbial interaction. We model mathematically mutualistic interactions within a minimal unit of structure: two growing reservoirs linked by a diffusive channel through which metabolites are exchanged. The model is applied to study a synthetic mutualism, experimentally parameterised on a model algal-bacterial co-culture. Analytical and numerical solutions of the model predict conditions for the successful establishment of remote mutualisms, and how this depends, often counterintutively, on diffusion geometry. We connect our findings to understanding 
complex behaviour in synthetic and naturally occurring microbial communities.
\end{abstract}

\pacs{87.23.Cc, 87.18.Hf, 87.10.Ca}
 
\maketitle

\section{\label{sec:intro}Introduction}

Microorganisms display a broad spectrum of interactions that determine the behaviour of 
microbial communities \citep{Abreu16}. Predicting this behaviour is a fundamental challenge in current microbial ecology \citep{Widder2016}. A wealth of experimental data on microbial community structure and dynamics is now available from `omics' approaches \citep{Cooper2015, Widder2016}. These, however, need to be complemented by lab-based studies of synthetic consortia and mathematical models to reach a mechanistic understanding of microbial dynamics \citep{Widder2016, Abreu16}. The study of mutualistic interactions between microbial populations is an active area of current research. Recent experimental studies have investigated synthetic mutualisms between microbes across the kingdoms of life. These include strains of enteric bacteria \citep{Harcombe2014, McCully2017, LaSarre2017} and yeast \citep{Allen2013} engineered to be mutualistic, and synthetic consortia combining wild type microbial species, such as bacterial tricultures \citep{Kim2008}, mixed cultures of algae and fungi \citep{Hom2014}, and algae and bacteria \citep{Croft2005, Kazamia2012a, Wang2014, Segev2016}.

Mutualistic interactions are conventionally modelled using Lotka-Volterra type models, with positive interaction coefficients \citep{Murray1989}. Linear mutualistic Lotka-Volterra models are known to display unrealistic unbounded growth \citep{Murray1989}, but logistic versions have been used to study demographically open mutualistic populations \citep{Thompson2006}, transitions between interspecies interactions \citep{Holland2009, Holland2010}, and the steady state dynamics of algal-bacterial co-cultures \citep{Grant2014a}. Since the pioneering work of May \cite{May1973}, such models have also been fruitfully employed to describe mutualistic interactions in network models of communities \citep{Okuyama2008}. In such models the interaction coefficients coupling species together define an interaction or community matrix (for mutualistic interactions the coefficients are positive and symmetric). Significant shortcomings of Lotka-Volterra models have recently been pointed out. For example, when species interact by exchanging metabolites, a metabolite-explicit model does not in general map onto a Lotka-Volterra implicit model \citep{Momeni2017}. Only in special instances does the microbial Lotka-Volterra form provide a good description of the microbial dynamics, e.g. when a fast equilibration approximation holds \citep{Hoek2016}. Resource-explicit models of bacterial mutualisms compare well with experiments in which mutualists are \textit{well-mixed} \citep{Lee1976,Wang2007, LaSarre2017, McCully2017}. Explicitly modelling resources is critical when studying spatially structured mutualistic systems (not well-mixed) whose interactions are controlled by metabolite dynamics and their spatial transport. 

Recent studies have considered spatial aspects of mutualistic and cooperative microbial interactions. Simulations using flux balance 
analysis (FBA) successfully predict the spatial growth on agar of colonies of synthetically mutualistic enteric 
bacteria \citep{Harcombe2014}. The FBA approach requires explicit knowledge of every known metabolic 
biochemical pathway in each mutualistic species, restricting its applicability to mutualisms between 
metabolically well-characterised organisms. Spatial effects on cheating \citep{Momeni2013a} and genetic 
drift \citep{Muller2014} observed in yeast colonies growing on agarose pads have also been modelled explicitly. In 
these models, coupled cells and nutrients diffusing in two dimensions are simulated to predict how 
nutrient-mediated interactions control spatial heterogeneity and survival of the populations. In general, 
interactions have been shown to control the spatial structure of laboratory biofilm communities \citep{Nadell2016}.
However, the homogeneous environment of nutrient agarose or laboratory biofilm substrates do not possess the intrinsic geometric or topological structure of natural microbial environments, such as the porous matrix of soil or microfluidic analogues \citep{Coyte2016}.
Mutualistic microbial dynamics have not thus far been studied in such structured environments, to the best of our knowledge.  

Here, we study a model of mutualistic microbial species in a simple geometry representing a minimal unit for a structured environment: populations growing in spatially separated reservoirs, 
metabolically linked by a channel. The model is generally applicable to auxotrophs cross-feeding remotely.
We apply it to make predictions for the dynamics of mutualistic populations of algae and bacteria diffusively exchanging vitamin B$_{\text{12}}$ and a carbon source, using model parameters obtained from independent co-culture experiments on this same mutualistic model system (see Appendix \ref{sup:appendix}). Such well-mixed co-cultures have been previously studied experimentally \citep{Kazamia2012a}. Our predictions provide new insights into the 
behaviour of microbial communities residing in structured geometries, both within synthetic consortia in the laboratory and environmental microbial communities. 

\section{Model}

\begin{figure}[tbh!]
\includegraphics[width=\linewidth]{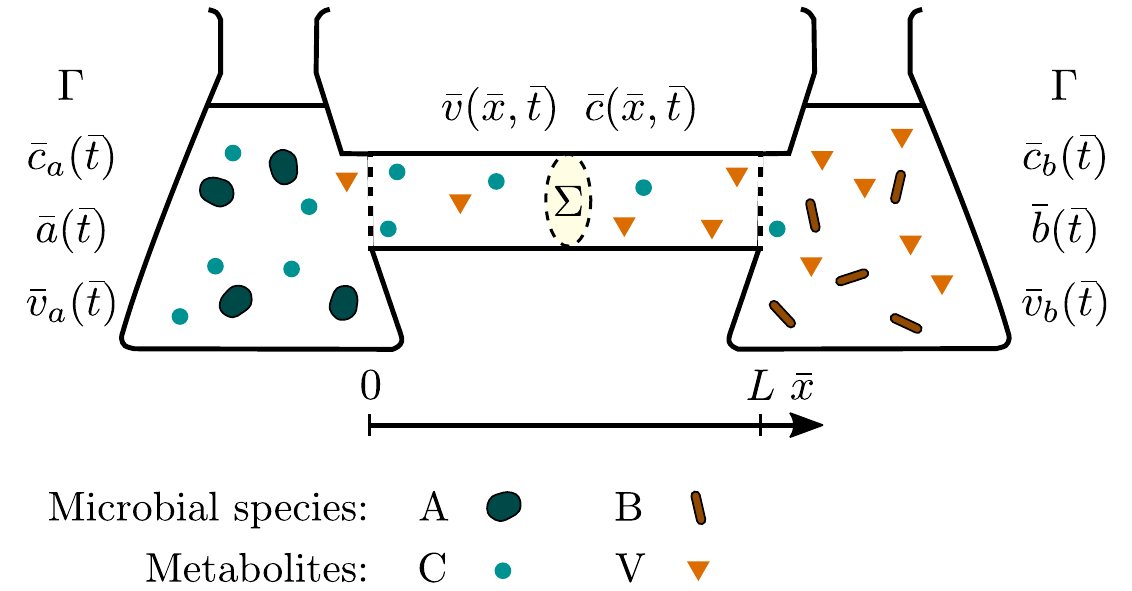}
\caption{
Diffusive cross-feeding at a distance. Auxotrophic microbial populations A and B (concentrations 
$\bar{a}$ and $\bar{b}$) reside in well-mixed reservoirs of equal volume $\Gamma$ separated by a channel of 
length $L$ and cross-section $\Sigma$. Microbe A produces a carbon source C, of homogeneous concentration $\bar{c}_a$, 
in its reservoir. This diffuses through the channel, forming a profile $\bar{c}(\bar{x},\bar{t})$, a function of position along the channel 
$\bar{x}$ and time $\bar{t}$. On reaching the reservoir where microbe B resides the concentration is homogenised to $\bar{c}_b$.
Symmetrically, the vitamin V produced by microbe B in its reservoir at concentration $\bar{v}_b$, diffuses to reservoir 
A creating a profile $\bar{v}(\bar{x},\bar{t})$, homogenised to $\bar{v}_a$ in the reservoir. Here, this general model is applied to an 
algal-bacterial partnership.
}
\label{fig1}
\end{figure}

The model describes two populations of mutualistic microbial species, A and B, interacting at a distance. 
The mutualistic interactions are predicated on auxotrophy: A requires metabolite V (for ``vitamin''), excreted by B; 
conversely B requires metabolite C (for ``carbon''), excreted by A. 
In formulating the problem we shall first use variables with an overbar to denote dimensional quantities (concentrations,
time, space), reserving symbols without typographical modification for appropriately rescaled variables.
Populations of A and B, with 
densities $\bar{a}(\bar{t})$ and $\bar{b}(\bar{t})$ 
respectively, reside in two well-mixed reservoirs, of equal volume $\Gamma$. These are spatially 
separated, but
connected by a cylindrical channel (length $L$, cross-sectional area $\Sigma$), as in Figure {\ref{fig1}}. 
The channel is impervious to cells, but porous to metabolite exchange by diffusion. Population A 
produces metabolite C with 
local concentration $\bar{c}_a(\bar{t})$, which diffuses out of the reservoir and into the channel at $\bar{x}=0$ (with $\bar{x}$ 
denoting the 
position along the channel axis), where it develops a spatial profile $\bar{c}(\bar{x},\bar{t})$ and eventually reaches the 
other reservoir 
at $\bar{x}=L$, where its concentration is $\bar{c}_b(\bar{t})$. Symmetrically, metabolite V produced by B with concentration 
$\bar{v}_b(\bar{t})$, 
diffuses out at $\bar{x}=L$ giving $\bar{v}(\bar{x},\bar{t})$, feeding the other reservoir at $\bar{x}=0$, generating a concentration $\bar{v}_a(\bar{t})$.

We first consider dynamics within the channel connecting the reservoirs, within which 
metabolites obey one-dimensional diffusion equations,
\begin{align}
\frac{\partial \bar{v}}{\partial \bar{t}} &= D_v \,\frac{\partial^2 \bar{v}}{\partial \bar{x}^2}~ & &\text{and} 
&\frac{\partial \bar{c}}{\partial \bar{t}} &= D_c \,\frac{\partial^2 \bar{c}}{\partial \bar{x}^2},
\label{eq:c_v_diff}
\end{align}
with $D_s$ the diffusion coefficients for metabolite S $=$ C or V. The boundary conditions to (\ref{eq:c_v_diff}) obtained 
from continuity 
at the channel-reservoir interface are: $\bar{c}_a(\bar{t})=\bar{c}(0,\bar{t})$, $\bar{c}_b(\bar{t}) = \bar{c}(L,\bar{t})$, 
$\bar{v}_a(\bar{t}) = \bar{v}(0,\bar{t})$, $\bar{v}_b(\bar{t}) = \bar{v}(L,\bar{t})$.
Clearly, one characteristic time scale of the problem is set by diffusive equilibration along the length of the channel,
\begin{equation}
 \tau_{\rm diff}=\frac{L^2}{D_s}~,
\end{equation}
where we anticipate that the diffusion constants of both metabolite species are similar.
From Fick's law, the flux $J_s$ (molecules area$^{-1}$ time$^{-1}$) of metabolite species S (C or V) entering, say, the left reservoir from the channel is 
\begin{equation}
J_s^0=D_s\frac{\partial \bar{s}}{\partial \bar{x}}\Bigg\vert_{0}.
\label{flux_define}
\end{equation}
The rate such molecules enter the reservoir is
$J_s^0\Sigma$, and with instantaneous homogenisation there, the rate of
change of the reservoir concentration $\bar{s}_a$ is $J_s^0\Sigma/\Gamma$.  The characteristic length
\begin{equation}
 \ell=\frac{\Gamma}{\Sigma}~
 \label{eq:ell_define}
\end{equation}
will play an important role in the model.  
If $\Delta \bar{s}$ is a typical difference in concentration of S between the two reservoirs, then the typical gradient
within the channel is $\Delta \bar{s}/L$, giving rise, by the arguments above, to an associated rate of change of reservoir 
concentration scaling as 
$d\bar{s}/d\bar{t} \sim (\Sigma/\Gamma) D_s\Delta \bar{s}/L \sim D_s \Delta \bar{s}/\ell L$, from which we 
can identify a characteristic 
equilibration time
\begin{equation}
\tau_\mathrm{eq}=\frac{\ell L}{D_s}~.
\label{eq:ratec_scaling}
\end{equation}
We define the ratio of equilibration and diffusive time scales to be
\begin{equation}
 \zeta\equiv \frac{\tau_{\rm diff}}{\tau_{\rm eq}}=\frac{L}{\ell}~.
 \label{zeta_define}
\end{equation}
The regime $\zeta\ll 1$ is that of fast establishment of the linear concentration
profile in the tube relative to changes of concentrations in the reservoirs, 
while for $\zeta \ge 1$ the transients
within the channel are on comparable time scales to that for changes in the
reservoirs. Semi-analytical solutions to the problem of chemical diffusion between two connected reservoirs further demonstrate the existence of these two regimes and the role of the previously identified timescales (see Appendix \ref{sup:appendix}).

We now turn to the population dynamics within the reservoirs, in which we
explicitly assume that algae reside in reservoir A and bacteria in B, and
that vitamin B$_{12}$ and carbon are exchanged.  
The dynamics obey the ordinary differential equations
\begin{widetext}
\begingroup
\addtolength{\jot}{1em}
\begin{subequations}
	\label{eq:LD_phy}
\begin{align}
   &\text{Reservoir A ($\bar{x}=0$)} &  &\text{Reservoir B ($\bar{x}=L$)} \notag\\
	\frac{d\bar{a}}{d\bar{t}}&=\mu_a \frac{\bar{v}_a}{K_v+\bar{v}_a} \bar{a} \left(1-\frac{\bar{a}}{K_a}\right) 
	- \delta_a \bar{a},                               
	& \frac{d\bar{b}}{d\bar{t}}&=\mu_b \frac{\bar{c}_b}{K_c+\bar{c}_b} \bar{b} \left(1-\frac{\bar{b}}{K_b}\right) 
	- \delta_b\ \bar{b},\label{eq:cells}\\
	\frac{d\bar{c}_a}{d\bar{t}}&=p_c \bar{a}+ \frac{1}{\ell}J_c^0, 
	&\frac{d\bar{c}_b}{d\bar{t}}&=-\mu_b\: \frac{\bar{c}_b}{K_c+\bar{c}_b}\; \frac{\bar{b}}{Y_b} 
	+ \frac{1}{\ell}J_c^L, \label{eq:carbon}\\
    \frac{d\bar{v}_a}{d\bar{t}}&=-\mu_a\: \frac{\bar{v}_a}{K_v+\bar{v}_a}\; \frac{\bar{a}}{Y_a}+\frac{1}{\ell}J_v^0, 
    &	\frac{d\bar{v}_b}{d\bar{t}}&=p_v \bar{b}+\frac{1}{\ell}J_v^L~, \label{eq:vitamin}
\end{align}
\end{subequations}
\endgroup
\end{widetext}
where $J_s^L = - D_s\frac{\partial \bar{s}}{\partial \bar{x}}\Bigg\vert_{L}$ is the flux of metabolite S = C or V entering the right reservoir.
In equations (\ref{eq:cells}) we model cell growth as logistic, with maximum growth rate $\mu_{i}$ and  
carrying capacity $K_{i}$ for species $i=$ A or B. Growth rates are limited by the abundance of the 
required metabolites. This is modelled using Monod factors \citep{Monod1949}, e.g., for C, 
$\mu_b \bar{c} / (K_c + \bar{c})$, where $K_c$ is the half-saturation constant (and symmetrically for V). 
Linear death terms, with mortality rates $\delta_{i}$ for $i=$ A or B, ensure exponential negative growth 
in the absence of the limiting metabolites. Equations (\ref{eq:carbon}) describe the dynamics of metabolite C. 
This is produced by species
A in proportion to its concentration with a rate $p_c$, and diffuses out at $0$. In the other reservoir, C is taken up 
by B. The uptake is assumed proportional to the cell growth rate, the proportionality constant is $1/Y_b$, 
where $Y_b$ is the yield coefficient (how much metabolite C results in a given concentration of species B).
Equations  (\ref{eq:vitamin}) describe the V dynamics, which are completely symmetric to the C dynamics.  Although inspired by bacterial-algal symbiosis, it is clear that
the structure of these dynamics is quite broadly applicable to mutualistic systems in general.

\subsection*{Identifying the key model parameters}

 In order to access the general dynamics of remotely cross-feeding monocultures, we nondimensionalise 
 equations \eqref{eq:LD_phy}. Because our focus is on the impact of geometry on 
 the biological processes, we choose a scheme accordingly.
 First, normalize the bacterial and algal concentrations by their respective carrying 
 capacities, the organic carbon and vitamin concentrations by their respective half-saturation concentrations, 
 rescale time by the bacterial growth rate, and rescale space by the length scale $\ell_b=\sqrt{D_c/\mu_b}$ 
 of organic carbon diffusion on
 the time scale of bacterial growth, defining
 \begin{eqnarray}
 a=\frac{\bar{a}}{K_a}, \ \ \ \ b&&=\frac{\bar{b}}{K_b}, \ \ \ \ c=\frac{\bar{c}}{K_c}, 
 \ \ \ \ v=\frac{\bar{v}}{K_v},\nonumber\\
  t&&=\mu_b\bar{t}, \ \ \ \ x=\frac{\bar{x}}{\ell_b}.
 \end{eqnarray}
The ratios of algal and bacterial growth rates and of their
 diffusion constants,  
 \begin{equation}
 \epsilon = \frac{\mu_a}{\mu_b}, \ \ \ \ \theta=\frac{D_c}{D_v},
 \end{equation}
 are two additional parameters.
 With now three characteristic lengths in the problem ($L,\ell, \ell_b$) one can form two independent dimensionless
 ratios.  These can be taken to be
 \begin{equation}
 \lambda=\frac{L}{\ell_{b}} \ \ \ \ {\rm and} \ \ \ \ \eta= \frac{\ell}{\ell_{b}},
 \end{equation}
 so that the parameter $\zeta$, defined previously in Eq. \ref{zeta_define}, is $\zeta=\lambda/\eta$.
 
 There are three pairs of parameters remaining which capture the relative strength of cellular death, 
 uptake and production in bacteria and algae respectively.  They are: the ratios of death rate to maximum growth rate of 
 bacteria and algae, which define mortality parameters
 	\begin{equation}
 	m_b = \frac{\delta_b}{\mu_b} \ \ \ \ {\rm and} \ \ \ \ m_a = \frac{\delta_a}{\mu_a},
 	\end{equation}
 	which must be less than $1$ for any population 
 	increase to occur; and finally, for both carbon and vitamin, the ratios of the typical uptake rate to the typical rate 
 	of change define the uptake parameters 
 	\begin{equation}
 	\kappa_b=\frac{K_b}{Y_b K_c} \ \ \ \ {\rm and} \ \ \ \ \kappa_a=\frac{K_a}{Y_a K_v};
 	\end{equation}
 	for both carbon and vitamin, the ratios of the typical production rate to the typical rate 
 	of change define the production strengths 
 	\begin{equation}
 	\sigma_c=\frac{p_c K_a}{\mu_b K_c} \ \ \ \ {\rm and} \ \ \ \ \sigma_v=\frac{p_v K_b}{\mu_a K_v}.
 	\end{equation}
 
With these rescalings, the dimensionless evolution equations are

\begin{widetext}
 \begingroup
 \addtolength{\jot}{0.5em}
 \begin{subequations}
 	\label{eq:LD_no_dim}
 	\begin{align}
 	\frac{1}{\epsilon}\frac{da}{dt}&= \frac{v_a}{1+v_a} a \left(1-a\right) - m_a a, & \frac{db}{d t}&= 
 	\frac{c_b}{1+c_b} b \left(1-b\right) - m_b b,  \label{eq:LD_no_dim_1}~\\
  \frac{d c_a}{d t}&=\sigma_c a + \frac{1}{\eta}j_c^0,  & \frac{d c_b}{dt}&=
  -\kappa_b\: \frac{c_b}{1+c_b}\; b - \frac{1}{\eta} j_c^{\lambda},\label{eq:LD_no_dim_2}\\
 \frac{1}{\epsilon} \frac{dv_a}{dt}&=
 -\kappa_a\: \frac{v_a}{1+v_a}\; a+\frac{1}{\epsilon \theta \eta} j_v^0,  
 & \frac{1}{\epsilon}\frac{d v_b}{d t}&=\sigma_v b - \frac{1}{\epsilon \theta \eta} j_v^{\lambda},
  	\label{eq:c_v_diff_no_dim}
 	\end{align}
 \end{subequations}
 \endgroup
 \end{widetext}
 
where now the dimensionless fluxes are $j_s^a=(\partial s/\partial x)_{x=a}$. These 
equations are to be solved together with the diffusion equations

\begin{equation}\label{eq:LD_no_dim_diff}
\frac{\partial v}{\partial t}=\frac{1}{\theta} \frac{\partial^2v}{\partial x^2} \ \ \ \ {\rm and} \ \ \ \
\frac{\partial c}{\partial t}=\frac{\partial^2c}{\partial x^2}
\end{equation}
 
for $c$ and $v$ on the interval $x \in [0,\lambda]$, ensuring continuity of fluxes and concentrations at the ends of the tube. Equations (\ref{eq:LD_no_dim}) were solved numerically to explore the role of diffusive geometry on mutualistic coexistence. We used the nondimensional parameters shown in Table \ref{tab:fit_phy_param}, corresponding to the mutualistic association between \textit{Lobomonas~rostrata}, a B$_{12}$-requiring green alga, and \textit{Mesorhizobium~loti}, a B$_{12}$-producing soil bacterium \citep{Kazamia2012a}. These parameter values were obtained by fitting growth and vitamin B$_{12}$ assay data (Figure~\ref{fig:fit_Vai}) from independent co-culture experiments we carried out with this model mutualistic system, as described in Appendix \ref{sup:appendix}. 

Before discussing the results from numerical solutions of the dynamical system of our model, we note that it supports a
trivial set of fixed points corresponding to reservoirs with no cells 
($a = b = 0$) and any combination of residual concentrations of metabolites. The non-trivial fixed point is given by
\begin{widetext}
\begingroup
\addtolength{\jot}{1em}
\begin{subequations}
	\label{eq:FP}
    \begin{align}
   a^{*}&=\frac{\sigma_c \sigma_v - \kappa_a \kappa_b m_a m_b}{\sigma_c (\sigma_v +\kappa_a m_a)},    &   b^{*}&
   =\frac{\sigma_c \sigma_v - \kappa_a \kappa_b m_a m_b}{\sigma_v (\sigma_c +\kappa_b m_b)}, \label{eq:FP1}\\
c_a^* &= c_b^{*} + \frac{\lambda \eta}{2} \left(\sigma_c a^* + \kappa_b \frac{c_b^{*}}{1+c_b^{*}} b^*\right), & c_b^{*}&
=\frac{\sigma_v (\sigma_c + \kappa_b m_b)}{(1-m_b) \kappa_b \sigma_v + \kappa_a m_a \kappa_b - \sigma_c \sigma_v}, \label{eq:FP2}\\
v_a^{*}&=\frac{\sigma_c (\sigma_v + \kappa_a m_a)}{(1-m_a) \kappa_a \sigma_c + \kappa_a m_b \kappa_b - \sigma_c \sigma_v}, & v_b^*&
=v_a^*+\frac{\lambda \epsilon \theta \eta}{2 } \left(\sigma_v b^* + \kappa_a \frac{v_a^*}{1+v_a^*} a^*\right).\label{eq:FP3}
    \end{align}
\end{subequations}
\endgroup
\end{widetext}

For the fixed point given by equations \eqref{eq:FP} to be physically relevant, the concentrations it describes must be positive. Therefore, the parameters must satisfy the following constraints:
\begingroup
\begin{subequations}
	\label{eq:FP_cond}
	\begin{align}
	& \sigma_c \sigma_v - \kappa_a \kappa_b m_a m_b >0~,\\
	& (1-m_b)\kappa_b \sigma_v+ \kappa_a m_a \kappa_b -\sigma_c \sigma_v >0~,\\
	\text{and}  \quad & (1-m_a)\kappa_a \sigma_c+ \kappa_a m_b \kappa_b -\sigma_c \sigma_v >0~.
	\end{align}
\end{subequations}
\endgroup
The first condition requires production strength to be strong enough to overcome cell mortality. This guarantees the existence of positive equilibrium 
algal and bacterial concentrations.  The second and third conditions guarantee this positivity for carbon and vitamin concentrations, respectively. They require that microbial consumption be high enough to overcome production. When these conditions are satisfied, the mutualistic microbes can reach a steady-state of co-existence. Note that in this steady-state, linear gradients of metabolite concentrations are present in the connecting tube.

\begin{table}[t!]
	\caption{\label{tab:fit_phy_param} Non-dimensional model parameters for the mutualistic association of \textit{M.~loti} and \textit{L.~rostrata} obtained from fitting independent co-culture experiments we carried out, as described in the text and Appendix \ref{sup:appendix}}.
	\setlength{\tabcolsep}{5pt}
	\smallskip
    \begin{ruledtabular}
	\begin{tabular}{l c c}
		Non-dimensional parameter                     &         Symbol          &   Value \\
        \hline
		\emph{Biological parameters}                         &                              &                \\
		Uptake parameter for algae                       &    \( \kappa_a \)    &      \num{1.3}       \\
		Uptake parameter for bacteria                  &     \(\kappa_b\)     &      \num{2.2}      \\
		Algal mortality/growth ratio                    &         \(m_a\)         &       \num{0.024}       \\
		Bacterial mortality/growth ratio               &         \(m_b\)           &     \num{0.014}       \\
		Carbon production strength                    &           \(s_c\)           &       \num{0.018}     \\
		Vitamin production strength                    &          \(s_v\)         &       \num{3.2}       \\
		Algal to bacterial growth rate ratio          &       \(\epsilon\)    &      \num{0.72}    \\ 
       \hline
		\multicolumn{2}{l}{\emph{Physical  parameters}}                                                  &                \\
		Ratio of metabolite diffusivities \footnote{obtained considering carbon with diffusivity \(D_c = \SI{5e-6}{\centi\meter\tothe{2}\per\second}\) as metabolite C and \BXII{} vitamin with diffusivity \(D_v = \SI{2e-6}{\centi\meter\tothe{2}\per\second} \) as metabolite V }                      &    \( \theta\)    &      \num{2.5}       \\
		Channel length                       &    \( \lambda\)    &      \numrange[range-phrase = --]{1}{30}       \\
		Equilibration length                      &    \( \eta \)    &      \numrange[range-phrase = --]{3}{100}       \\
	\end{tabular}
    \end{ruledtabular}
\end{table}

\subsection*{Feeding on a distant passive source}

Before considering the fully coupled system dynamics, we consider the case of a single auxotrophic species B, concentration $b$, residing in a reservoir initially 
free of a growth-limiting metabolite coupled by the channel (also initially nutrient-free) to a strong source 
of the metabolite with initial concentration $c^0_a$. This source consists of a reservoir filled with limiting metabolite. 
The long time steady-state for the model is always extinction of B once it has exhausted the remote resource. 
However, separation of the microbial population from the source modifies the transient population dynamics.  Recalling the nondimensional channel length $\lambda=L/\ell_{b}$ 
and equilibration length $\eta=\ell/\ell_{b}$, 
we can define the nondimensional timescales $t_\mathrm{diff}=\lambda^2$ and $t_\mathrm{eq}=\lambda \eta$ 
as the ratios between the typical times of diffusion and of equilibration between reservoirs, and 
the
biological growth timescale $\tau_b = 1/\mu_b$. These ratios gauge the 
relative rates of diffusion/equilibration and growth. We require $t_\mathrm{eq}$ and 
$t_\mathrm{diff} \sim 1$ for 
diffusion to transport metabolites to species B, stimulating its growth.

\begin{figure}[tbh!]
	\includegraphics[width=0.9\linewidth]{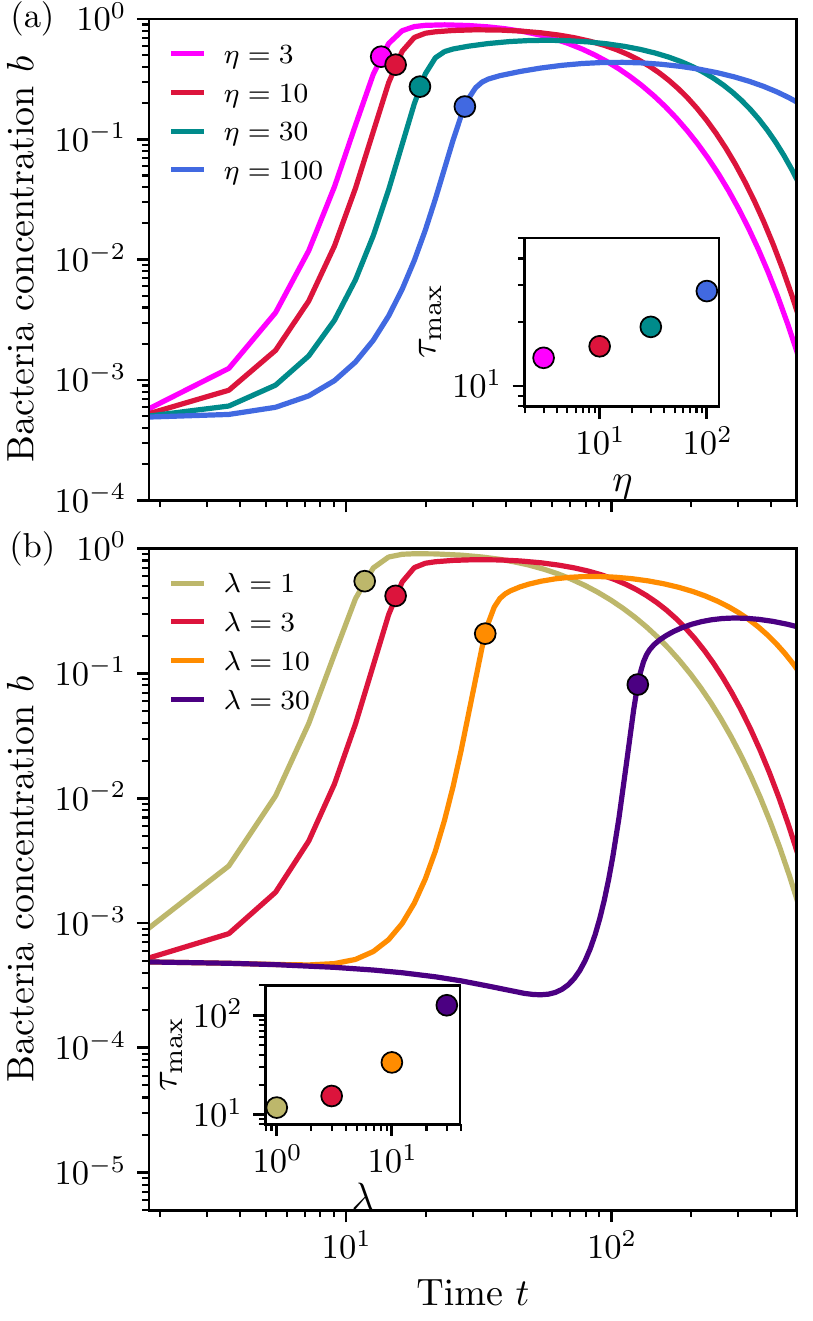}
	\caption{Transient dynamics of a bacterial population fed through a channel that allows metabolite diffusion 
from a remote carbon source. The diffusive exchange geometry controls the dynamics through the nondimensional 
channel length $\lambda$ and reservoir equilibration length $\eta$. Model solutions predict that: (a) for fixed 
$\lambda = 3$, increasing $\eta$ delays the time of peak bacterial growth and curtails growth due to a limited 
carbon-source flux; (b) for fixed $\eta = 10$, increasing $\lambda$ significantly delays peak growth, with an 
impact on the maximum bacterial concentration attained. The delay as measured by $\tau_\mathrm{max}$, the 
time of maximal growth rate, is proportional to $\lambda^2$ (inset). For all simulations, initial nondimensional 
bacterial and carbon concentration are $b_0 = \num{5e-4}$ and $c_a(t=0) = 10$, other parameters are from 
table~\ref{tab:fit_phy_param}.}
	\label{fig:remote_monoculture}
\end{figure}

We have solved the remotely-fed single microbe limit of the model numerically (see Appendix \ref{sup:appendix}) to predict the 
dynamics of the rhizobial bacterium \textit{Mesorhizobium loti} fed from a remote glycerol carbon source. 
Figure \ref{fig:remote_monoculture} shows the transient growth dynamics in the regime for which both geometric parameters $\lambda$ and $\eta$ impact the dynamics. 
We first consider the effect of diffusive reservoir equilibration, quantified by $\eta$ for a fixed channel length $\lambda$. 
For large $\eta$, $t_\mathrm{eq}$ is large: diffusive equilibration in the reservoir is much slower than growth. Thus, 
the instantaneous flux from the carbon source reservoir to the bacterial reservoir is below what the bacteria need to 
grow to carrying capacity. As a result, increasing $\eta$ decreases the value of the peak bacterial concentration 
(preceding the inevitable decay), as well as delaying the onset of growth (Figure \ref{fig:remote_monoculture}a).  
Next we fix $\eta$ and vary $\lambda$. Since the diffusive timescale scales like $\lambda^2$, increasing $\lambda$ 
progressively delays the onset of bacterial growth (Figure \ref{fig:remote_monoculture}b inset). Large $\lambda$ 
values also correspond to weaker carbon source gradients across the tube, and thus a `slow-release' nutrient flux.  
Consequently, a less concentrated population can be sustained for longer by the remote source 
(Figure \ref{fig:remote_monoculture}b). The passive source case we have just considered demonstrates the critical role played by both geometric parameters $\lambda$ and $\eta$ in setting the timescale of transients, but also the peak microbial numbers achievable on a finite resource. 

\subsection*{Remotely cross-feeding populations}

\begin{figure*}[tbh!]
	\includegraphics[width=0.85\linewidth]{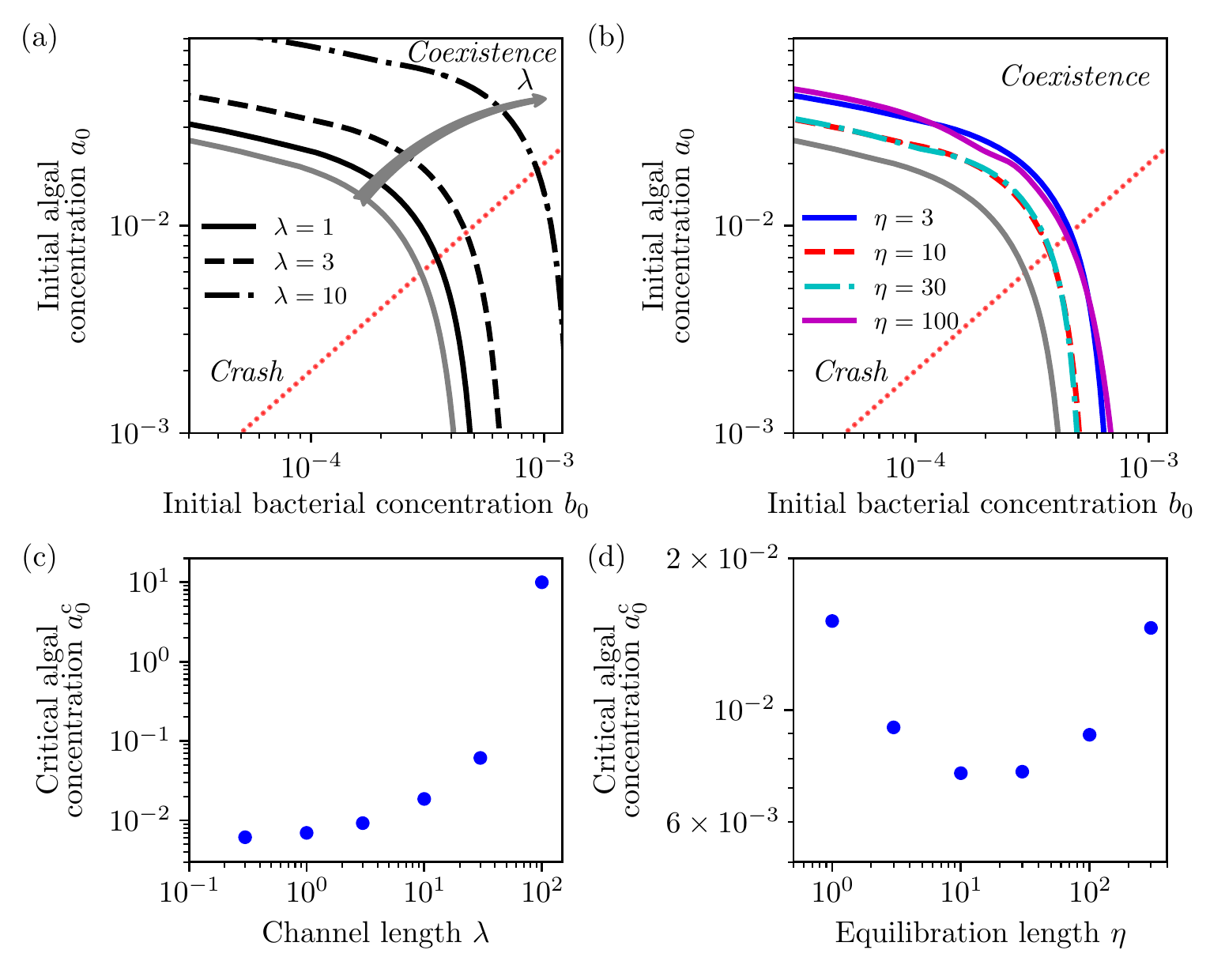}
	\caption{Coexistence diagram illustrating the long-time fate of mutualistic populations in terms of initial concentrations. (a) At a fixed equilibration length $\eta = 3$, increasing channel 
	length $\lambda$ causes the coexistence region to shrink progressively. (b) On the other hand, the response to an 
	increase in $\eta$ for fixed channel length $\lambda = 3$ is nonmonotonic. The coexistence initially contracts, 
	then expands, and finally contracts again. The grey lines in both plots corresponds to the membrane limit for 
	which $\lambda \to 0$ and equilibration of metabolite concentrations between the two flasks is instantaneous. This provides
	the maximum possible concentration parameter space for mutualistic coexistence. The coexistence boundaries were 
	determined by solving equations \eqref{eq:LD_no_dim} and \eqref{eq:LD_no_dim_diff} numerically using the parameters in 
	table~\ref{tab:fit_phy_param} (see Appendix \ref{sup:appendix}). (c) Along the transect (dotted red line) in (a) corresponding to a conserved 
	ratio of initial concentrations $b_0/a_0 = 20.0$, the critical initial algal concentration $a_0^c$ above which coexistence 
	occurs is an increasing monotonic function of the length of tube $\lambda$. (d) Using the same transect in (b), 
	the non-monotonic behavior of the critical algal concentration $a_0^c$ with $\eta$ is clearly revealed.}
	\label{fig:surv_map2}
\end{figure*}

Next, we consider auxotrophic populations in separate reservoirs, exchanging limiting metabolites through a connecting channel. 
As mentioned earlier, we apply the model to an algal-bacterial system, obtaining our parameters from experiments where the phototrophic alga 
{\it L. rostrata}, auxotrophic for vitamin B$_{\rm 12}$, is grown in co-culture with the heterotrophic 
bacterium \textit{M. loti}. The algal and bacterial populations in their reservoirs 
have initial concentrations, $a_0$ and $b_0$, respectively.  Neither carbon source nor vitamin (the limiting metabolites) are 
initially present in the reservoirs and channel. The coexistence diagrams in Figure \ref{fig:surv_map2}a,b show what values in 
the initial concentration parameter space give rise to long-term mutualistic coexistence or a population crash due to metabolite deprivation. These fates are the possible fixed points of our model, which we shall also refer to as model equilibria. 
Figure \ref{fig:surv_map2}a displays the boundary between these two regions for different values of the channel length 
$\lambda$ for a fixed value of the equilibration length $\eta$. 
In Figure \ref{fig:surv_map2}b crash-coexistence boundaries are instead shown for different equilibration lengths 
$\eta$ at fixed $\lambda$. Also shown on both diagrams is the \textit{membrane limit} (bottom-left grey line). In this limit the 
distance between reservoirs vanishes ($\lambda \to 0$) and they are simply separated by a membrane impervious to cells, 
as has been demonstrated experimentally in co-culturing/metabolomic experiments \citep{Paul2013}. We assume instantaneous equilibration of metabolite concentrations across the membrane in this limit. It is thus an ideal case in which exchanges are not limited by diffusion dynamics along the tube nor by the geometry of the problem, and as such represents an interesting common reference case to understand the impact of both the channel length $\lambda$ and the equilibration length $\eta$.

We see that increasing the channel length has the effect of pushing the crash-coexistence boundary toward higher initial microbial concentrations (Figure~\ref{fig:surv_map2}a,c). Coexistence is achieved in the membrane limit 
for initial concentrations lower than those for finite $\lambda$. The boundary between crash and coexistence regions 
shifts quantitatively with $\lambda$, but does not change significantly qualitatively. Its shape is revealing: if the initial 
concentration of bacteria $b_0$ is not too large, coexistence depends weakly on $b_0$, and very strongly on the initial 
algal concentration $a_0$. For low enough bacterial concentrations, the smallest critical initial algal concentration for 
which coexistence will occur increases with $\lambda$. These features are reasonable considering that there is a diffusive 
delay in the metabolite exchange between reservoirs: if the delay is too long, auxotrophs will difficultly recover in the 
absence of a limiting nutrient. However, we note that the model does not predict any critical length above which recovery is impossible: longer separations will simply restrict the establishment of co-existence to cases with very high initial populations.

The effect of the reservoir equilibration length $\eta$ on the coexistence diagrams is more subtle. Recall $\eta$ is the nondimensional ratio of growing volume to metabolite exchange area, which controls diffusive equilibration in the reservoirs. For small $\eta$, the crash-coexistence boundary sits above the membrane limit boundary toward higher initial concentrations. This boundary is then pushed toward lower initial concentrations for intermediate values of $\eta$ while still sitting above the membrane limit (as expected given that the membrane limit corresponds to the ideal case of instantaneous equilibration for no separation length), before raising to higher initial values for high values of $\eta$ (Figure~\ref{fig:surv_map2}b,d).
The general shape of the boundary is preserved for all $\eta$. 
To understand the nonmonotonic dependence of the boundary shift with $\eta$, we note $\eta/\lambda$ is the 
reservoir/channel volume ratio.
Thus, with $\lambda$ fixed, changing $\eta$ takes the populations through three regimes: i) the reservoir volume is 
small compared to that of the channel, $\eta/\lambda\ll1$; ii) the volumes are the same size, $\eta/\lambda\sim 1$; 
iii) the channel volume is smaller than that of the reservoir, $\eta/\lambda\gg1$. In regime i), the equilibration time  
$t_\mathrm{eq}=\lambda \eta$ is small, but a large channel volume relative to the reservoirs dilutes any metabolite produced, making metabolites inaccessible to the microbial partner and preventing co-existence. In regime iii), the 
relative channel volume is small, but co-existence is impeded due to the 
long equilibration time $t_\mathrm{eq}\gg1$, which slows down significant metabolite exchanges between reservoirs.  
Finally, in regime ii), where reservoirs and channel have similar volume and $t_\mathrm{eq}\sim 1$, mutualistic 
coexistence is favoured.

\begin{figure}[tbh!]
	\includegraphics[width=0.9\linewidth]{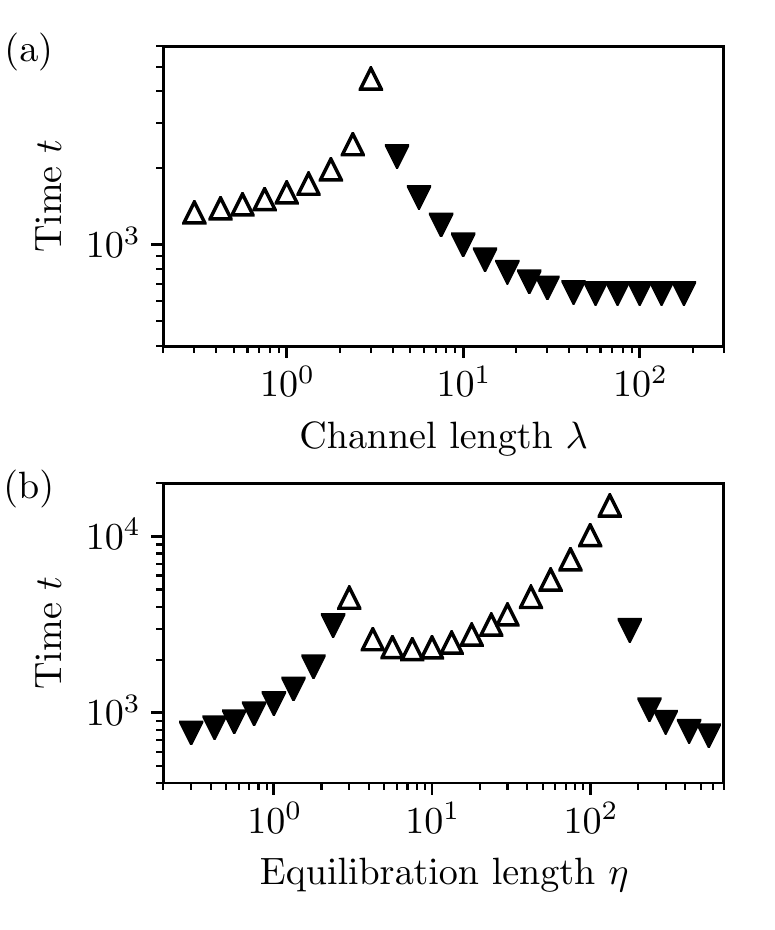}
	\caption{The time taken for the populations to relax to equilibrium (crash or coexistence) depends on the geometric 
	parameters $\lambda$ and $\eta$. Here, we plot these times for fixed initial microbial concentrations 
	$(a_0, b_0)$ (assuming, as before, no initial metabolites within the diffusion geometry). Times for populations reaching 
	coexistence are shown in white up-pointing triangles, and those for populations that will crash in black down-pointing triangles. (a) For fixed $\eta=3$, 
	the relaxation time increases with $\lambda$ up to the critical value at the coexistence boundary (where it diverges).  
	On the other side of this critical value it decreases. (b) For fixed $\lambda=3$, the dependence of the time as a function of 
	$\eta$ shows a similar divergence when approaching a transition between extinction and survival. For the initial 
	concentrations $(a_0, b_0)$ here chosen, two of these transitions are possible, with extinction for low and high values of 
	$\eta$ and coexistence for intermediate values. Both panels correspond to $a_0 = \num{2e-2}$ and $b_0 = \num{3e-4}$.}
	\label{fig:time2SS}
 \end{figure}

Aside from the co-existence or crash fixed points just discussed, we can use the model to analyse the transient dynamics 
leading to these equilibria. In particular, it is illuminating to evaluate the relaxation time taken for remote populations to 
reach the fixed points for a given initial microbial concentration in reservoirs assumed initially devoid of metabolites, as 
previously. Numerical solutions of the model equations show that this time varies as $\lambda$ is increased across 
the co-existence/crash boundary for given $\eta$, as shown in Figure \ref{fig:time2SS}a. It is clear that the time to 
relax to the equilibrium rises sharply on either side of the critical $\lambda$ at the boundary. This slow relaxation for 
$\lambda$ values close to the bifurcation between extinction or co-existence is accompanied by oscillatory transients (see Figure \ref{fig:time2SS_sup}). Similar considerations apply to the dependence of this time on the equilibration length 
$\eta$ for a given $\lambda$, within that case there is the possibility of two boundaries between extinction and survival, see 
figure \ref{fig:time2SS}b. We thus predict a complex behavior of the time needed to reach steady-state in such connected mutualistic systems, with the potential for slow relaxation if geometrical parameters are close to critical values between extinction and co-existence.

Interestingly, the algal and bacterial concentration fixed points, $a^*, b^*$ respectively, are independent of 
$\lambda$ and $\eta$, as already mentioned (see equations \eqref{eq:FP}). Larger separation (increasing $\lambda$) or weaker diffusive 
coupling to the reservoirs (increasing $\eta$) increases delays in chemical exchanges and reduces the extent of the mutualistic 
co-existence region. However, these geometric changes do not alter the microbial concentration fixed points, which have 
the same values as in the membrane limit: high densities of mutualistic microbes can be achieved even with weak or slow diffusive coupling.
This equilibration is possible thanks to supply of metabolites (whose concentrations are also geometry-independent, see 
equations \eqref{eq:FP}) from the partner reservoir. A sufficiently large metabolite gradient across the channel is required 
to support the equilibrium metabolite and cell concentrations. Indeed, the model predicts an increase in the metabolite 
concentration at the production reservoir.  For example, if the equilibrium concentration of vitamin B$_{12}$ in the algal reservoir is $v_a^*$, then at the bacterial reservoir we predict $v_b^* = v_a^* + \lambda \eta f(a^*, b^*, v_a^*)$, where the function 
$f$ can be obtained by comparison with equation \eqref{eq:FP}. The same applies for carbon. This metabolite enrichment is an interesting prediction of the model. The concentration excess at the 
production reservoir is linear in both separation $\lambda$ and equilibration length $\eta$: two parameters with which 
enrichment could be experimentally controlled. As an example, for the \textit{L.~rostrata} and \textit{M.~loti} mutualism
using $\lambda=1.25$ and $\eta=2$ (all other parameters as before) our model predicts a sevenfold enrichment of 
vitamin B$_{12}$ in the bacterial reservoir compared to the algal side. 

\section{Discussion}

Microbial populations often interact by diffusive exchange of metabolites in structured environments, such as the porous matrix of soil. Metabolite diffusion is known to play an important role in determining microbial dynamics in unstructured environments \citep{Allen2013,Momeni2013a,Harcombe2014,Hom2014,Nadell2016}. Current models of microbial interactions, however, do not explicitly model diffusive transport in geometrically confining habitats. A recent theoretical study has investigated microbial invasion in soil networks \citep{Perez-Reche2012}, but interactions were modelled stochastically, without considering diffusive exchanges. How the geometry of diffusive exchanges constrains microbial interactions remains an important open question. We have addressed this here by modelling a minimal geometrical unit of microbial interaction: two mutualistic populations in finite volume reservoirs linked by a diffusive channel. The model was solved to predict the diffusively mediated interactions of mutualitistic algae and bacteria, whose dynamics in co-culture have been experimentally characterised \citep{Kazamia2012a}. Two key geometrical parameters control the diffusive exchange of metabolites between the populations: the separation $\lambda$ (the nondimensional channel length) and the equilibration length $\eta$ (the nondimensional ratio of growing volume to metabolite exchange area). Model solutions allow prediction of whether initial concentrations of algae and bacteria will result in mutualistic coexistence or population crash (the model equilibria) for given values of the geometrical parameters $\lambda$ and $\eta$. In particular, we can draw the boundary between regions exhibiting these two equilibria for given initial microbial concentration, and predict how this boundary shifts when the values of the geometrical parameters are changed.

The model makes several interesting predictions. For instance, coexistence between mutualistic partners can be achieved only if the numbers of one or both partners are abundant; low initial numbers will lead to a crash. This feature is
qualitatively independent of diffusive geometry ($\lambda$ or $\eta$), like the shape of the coexistence boundary itself (approximately flat for a broad range of bacterial concentrations, falling very rapidly thereafter, see Figure \ref{fig:surv_map2}). It has an intuitive explanation: an initially high concentration of one of the two species will produce a large initial amount of metabolite, which allows the partner species to grow and recover, even from initially very low numbers. A more surprising result is that mutualistic populations at a distance can achieve as high a steady concentration as in a mixed environment. The effect of the diffusive geometry is only to modify the transient dynamics and raise the initial cell concentration values required to avoid a crash. The fact that, given enough time, separated cross-feeding mutualists might reach as high numbers as populations in proximity is a counterintuitive result of great potential significance for microbial ecology. This contrasts with the case of a population feeding from a distant passive resource (Figure \ref{fig:remote_monoculture}), for which maximum achievable concentrations do depend strongly on geometric coupling. 

A final prediction of the model to highlight is the nonmonotonic dependence of the boundary position as the equilibration length $\eta$ is varied. As one might expect, increasing the channel length $\lambda$ (at fixed equilibration length $\eta$ and bacterial concentration $b_0$) increases the critical concentration of algae that will support co-existence with bacteria. On the other hand (for fixed $\lambda$ and $b_0$) the critical algal concentration varies nonmonotonically, falling and then rising again with increasing $\eta$. The dependence on $\lambda$ is intuitive: separating the partners further increases a diffusive delay, 
which we recall scales like $\lambda^2$, so that more algae are required to support coexistence at a distance.  
The nonmonotonic behaviour with $\eta$ is less obvious. It results from a dilution of metabolites in the volume of the 
channel for low values of $\eta$, requiring higher initial densities for successful coexistence, and from weak fluxes of 
metabolites into the homogenisation volume when $\eta$ is large. With respect to these two extremes, coexistence is more easily achieved at intermediate values of $\eta$. This is another counterintuitive prediction, which highlights the value of explicitly accounting for diffusive transport in modelling mutualistic interactions.

Our findings have implications for the microbial ecology of synthetic consortia. This is an active area of investigation, with several recent studies on microbial mutualisms \citep{Kim2008,Kazamia2012a, Allen2013, Harcombe2014, Hom2014,Wang2014,Segev2016,McCully2017,LaSarre2017}. None thus far have addressed the role of diffusive geometry on these interactions, which could test the predictions of our model. A preliminary experiment in which batch cultures of algae and bacteria grow linked by a channel allowing metabolite diffusion (filled with a hydrogel to prevent cross-contamination) demonstrates the possibility of establishing remote mutualisms, see Appendix \ref{sup:appendix}. Further, it provides preliminary confirmation that vitamins accumulate in the B$_{12}$ producer (bacteria) flask, as predicted by our model (equation \eqref{eq:FP}). The experiment provides a `proof of concept' and a blueprint for further experiments using our connected flasks set-up. These should explore how the population behaviour varies with the geometrical parameters, and if the stark predictions of the model, such as the nonmonotonicity of the crash-coexistence boundary with $\eta$, are borne out experimentally. Alternatively, experiments using diffusively coupled microfluidic chambers \citep{Kim2008,Karzbrun2014}, could be used, noting that modifications would be necessary to account for stochastic effects associated with the small cell numbers in such systems \citep{Khatri2012}. As well as  being tested, the model could be used to describe other 
synthetic consortia in which populations also interact diffusively across porous hydrogels \citep{Kazamia2012a, Harcombe2014} or microfluidic structures \citep{Kim2008}. It is straightforward to extend the model to account for two or three-dimensional diffusive exchanges appropriate to these systems.

\begin{figure}[tbh!]
	\includegraphics[width=\linewidth]{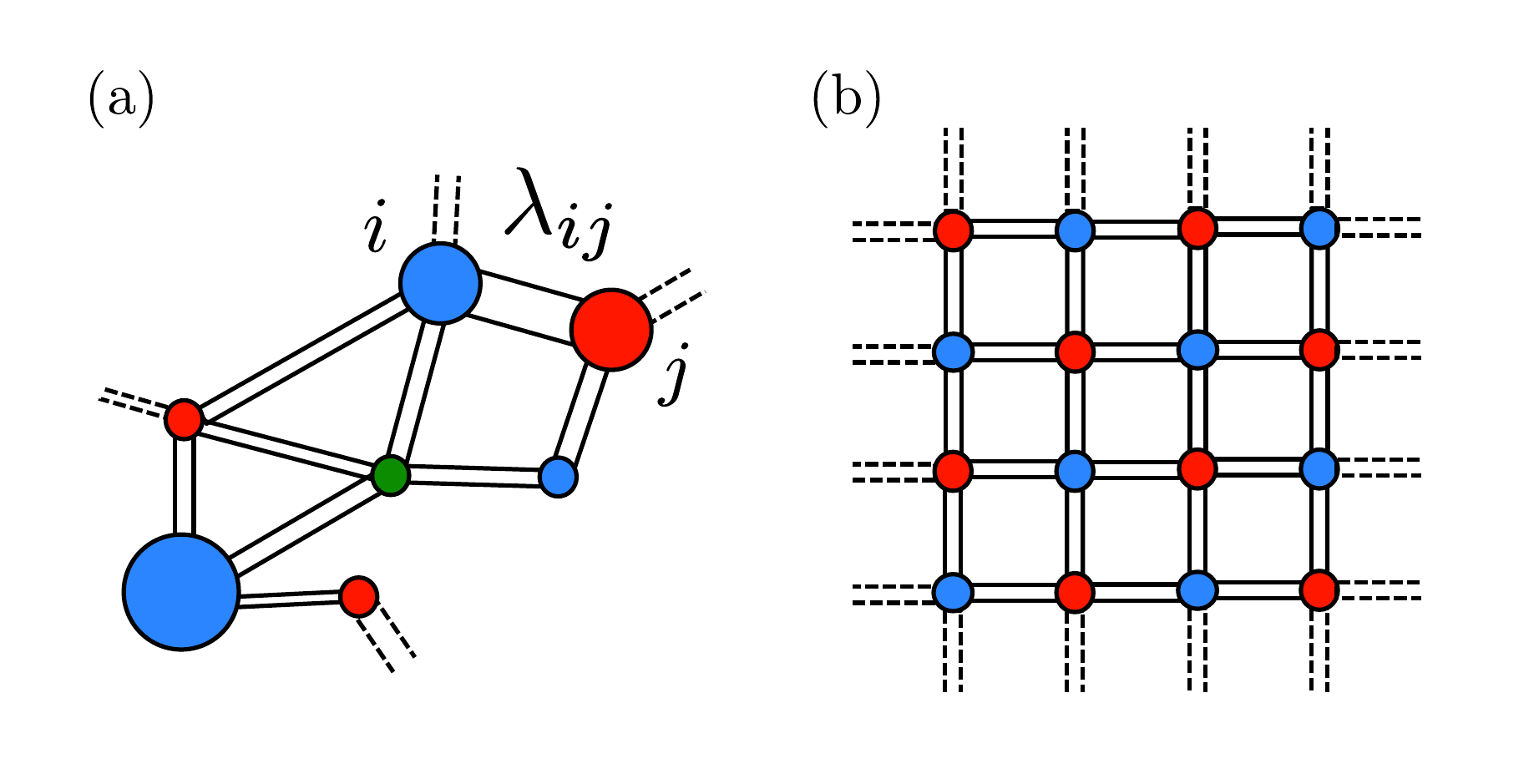}
	\caption{
    Schematic of a diffusively coupled microbial network representing: (a) A 
    structurally and microbially heterogeneous network as a realistic representation of soil \citep{Perez-Reche2012}; (b) A crystalline network that can be engineered in the laboratory. The nodes of this physically structured network represent reservoirs of different volumes filled with different growing microbial species diffusively exchanging metabolites via porous channels, as described in the model formulated in this work. Diffusive exchanges are parameterised by sets of geometric parameters, as such as the lengths, $\lambda_{ij}$, of the channels connecting nodes.} 
	\label{fig:soilnetwork}
\end{figure}

The present model may also provide the foundation for a physical description of microbial networks, e.g. 
consortia for cooperative biosynthesis \citep{Hom2015, Cavaliere2017} or microbial communities in soil, or spatially coupled biofilms \citep{Liu2017}. Indeed, as mentioned earlier, at the microbial scale, 
soil can be approximated as a physical network of growth chambers linked by channels \citep{Perez-Reche2012}. In establishing the key geometric parameters that govern the most elementary unit in a network, namely
two diffusively linked nodes (reservoirs), the present work provides a basis for describing population dynamics in a two- or three-dimensional network
of coupled nodes (Figure~\ref{fig:soilnetwork}). It is left to future work to take up the significant 
challenge of studying such networks, particularly when there is inhomogeneity in the diffusive couplings and stochasticity in the populations themselves. This view of microbial networks centering on the physics of diffusion could also help refine interaction 
matrix models of microbial communities and extend them beyond contact interactions \citep{Mathiesen2011}. An interesting possibility is that interaction networks could be simplified by constraints deriving from diffusion geometry. 

Aside from the microbial networks mentioned above, 
the model may also be a relevant interpretative tool to understand the behaviour of structured environmental 
communities with diffusive exchanges, such as river biofilms \citep{Battin2016} or sediment layers \citep{Pagaling2014}. Moreover, knowledge of the mechanisms for metabolite exchange between spatially separated organisms is important to gain insight into how such communities initiate in the natural environment, and the drivers and constraints on the evolution of mutualisms within them \citep{Kazamia2016}.

\subsection*{Acknowledgments}

We thank J. Kotar and R. Bowman for discussions. We thank the Cavendish and G. K. Batchelor Laboratory workshops for assistance, in particular D. Page-Croft. F.J. Peaudecerf gratefully acknowledges support from Mines ParisTech and from a Raymond and Beverly Sackler Scholarship. O.A. Croze, M.A. Bees and A.G. Smith gratefully acknowledge support from the Engineering and Physical Sciences Research Council (EP/J004847/1). O.A. Croze also acknowledges support from a Royal Society Research Grant and the Winton Programme for the Physics of Sustainability. R.E. Goldstein  acknowledges support from an EPSRC Established Career Fellowship (EP/M017982/1) and the Schlumberger Chair Fund. V. Bhardwaj was in receipt of a studentship from the Gates Cambridge Trust. F. Bunbury is in receipt of a studentship from the UK Biotechnology and Biological Sciences Research Council (BBSRC) Doctoral Training Partnership.

\newpage
\appendix
\renewcommand*\thetable{\Alph{section}.\arabic{table}}
\setcounter{table}{0} 

\section{}\label{sup:appendix}

\subsection{Diffusive reservoir equilibration (no microbes)  \label{sup:connected_reservoir}}

We consider here the purely physical equilibration between two diffusively connected reservoirs to reveal the interplay between the diffusive time and the equilibration time in such a system.  This setup utilises the same 
geometry as in Fig. \ref{fig1}, with the reservoir at $\bar{x}=0$ having an
initial
concentration $\bar{c}_0 (\bar{t} = 0) = \bar{c}_{\text{init}}$ of a chemical species, and the reservoir at $\bar{x}=L$ having an initial concentration $\bar{c}_L(t=0) = 0$ of the same species.  
The chemical concentration along the tube is initially equal to zero, and has 
diffusivity \(D\). 
Since our focus here is purely on the different physical timescales independent of biological processes, we choose a non-dimensionalisation scheme restricted to this section only that differs from the main body of the paper. Rescaling chemical concentrations by $c_\text{init}$, 
lengths by $L$ and time by $L^2/D$, we obtain 
\begin{eqnarray}
    \label{eq:num_nodim}
		\frac{\partial c}{\partial t} ~=&&  \frac{\partial^2 c}{\partial x^2}~, \nonumber\\
		\frac{d c_0}{d t} = \zeta \left.\frac{\partial c}{\partial x} \right|_{x=0}~, &&~
		\frac{d c_L}{d t} = - \zeta \left.\frac{\partial c}{\partial x} \right|_{x=1}~, 
\end{eqnarray}
where we recognise the nondimensional parameter $\zeta = L/\ell$, the ratio of tube length $L$ to equilibration length $\ell = \Gamma/\Sigma$. These equations are subject to initial conditions $c_0(0) = 1~, \;  c_L(0)=0~, \;   c(x,0) = 0$ 
and boundary conditions $c_0(t) = c(0,t)~\mathrm{and} \; c_L(t) = c(1,t)$.
Despite the fact that this is a linear PDE with apparently simple boundary
conditions, the fact that it exists on a finite domain, and is coupled
to the reservoir dynamics, makes it difficult
to obtain an explicit analytical solution for general values of $\zeta$.

\subsubsection{Approximate solution for \(\zeta \ll 1\)}

When  $\zeta \ll 1$, the time evolution of the reservoir concentrations is much slower than the establishment of a concentration gradient in the tube.  
Thus, the diffusive dynamics within the tube reach a quasi-steady-state distribution between the two reservoir concentrations $c_0(t)$ and $c_L(t)$.  
In this approximation, the solution to the diffusion equation in the tube is 
the linear profile $c(x,t) \approx \left[c_L(t) - c_0(t)\right] x~.$
Substituting this solution into the reservoir dynamics, and solving the resulting two ODEs yields (in dimensional units)
\begin{equation}
	\bar{c}_L(t) \approx \frac{\bar{c}_{\text{init}}}{2}\left[1-\exp(- \bar{t}/\tau_\text{eq})\right]~.
	\label{eq:exp_cl}
\end{equation}
We thus deduce that in the limit $\zeta = L/\ell \ll 1$, the timescale of exchanges is purely dominated by the equilibration time $\tau_{\text{eq}} = L \ell / 2 D$, as argued previously.  
The same time scale plays a role when the biological dynamics of growth and production are considered, as discussed in the main text. 

\begin{figure*}[tbh!] 
	\includegraphics[width=0.9\textwidth]{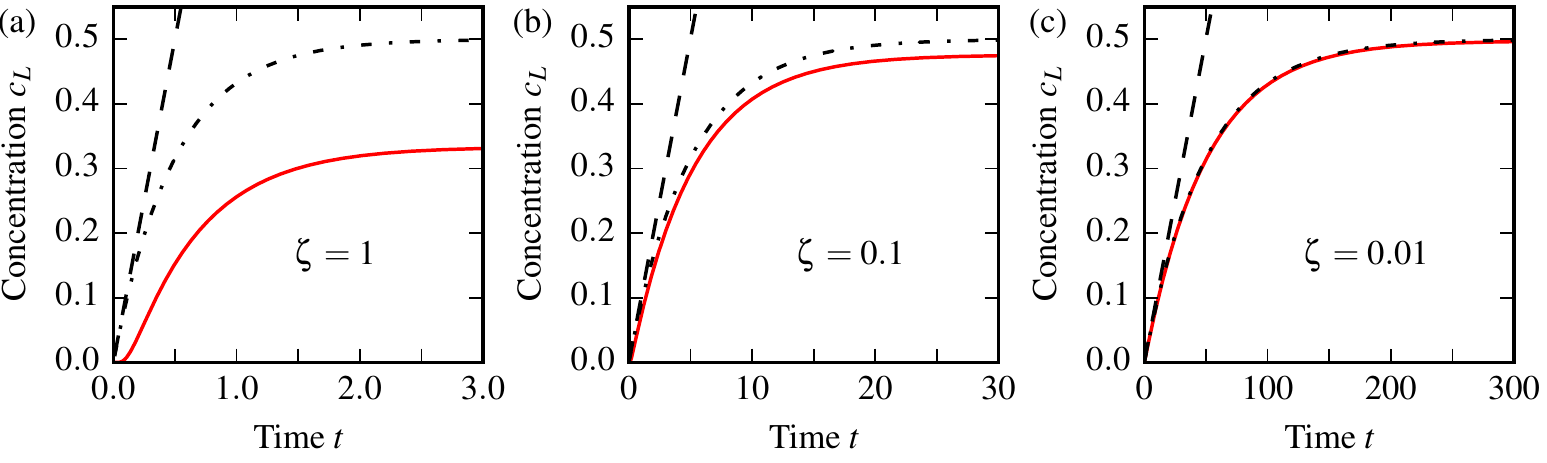}
	\caption{Evolution of the concentration $c_L$, in a reservoir initially devoid of chemical, diffusively coupled to a reservoir filled with initial concentration 
	$c_0(0) = 1$. The concentration was evaluated numerically from the inverse Laplace transform of $f_L$, given in equation \eqref{eq:fL}. Each red curve in panels (a), (b) and (c) 
	corresponds to a numerical evaluation for  value of the parameter $\zeta$ equals to $1$, $0.1$ and $0.01$ respectively. Dash-dotted lines are the corresponding nondimensional versions 
	of the approximation of $c_L$ as a saturating exponential as given in equation \eqref{eq:exp_cl}, while dashed lines correspond to the linear approximation $c_L = \zeta t$. Note the change of scale of the time axis for 
	different values of $\zeta$, where time itself has been rescaled
    by $L^2/D$.}
	\label{fig:res_diff}
\end{figure*}

\subsubsection{General solution from Laplace transform}

To find the general solution of this problem, we examine the Laplace transforms of the nondimensional concentrations 
$\Lapl(c_0)(s) = f_0 (s)$, $\Lapl(c_L)(s) = f_L (s)$, and $\Lapl(c)(x, s) = f (x, s)$.
Laplace transforming the diffusion equation in the tube we find the general
solution
\begin{equation}
	f(x,s) = M(s)\exp(x \sqrt{s} ) + N(s) \exp(-x \sqrt{s} )
\end{equation}
with $M(s)$ and $N(s)$ functions of the Laplace variable to be determined.
Imposing boundary conditions at the tube ends gives
\begin{subequations}
	\label{eq:l1}
	\begin{align}
		f_0(s) &= M(s) + N(s)\\
		\text{and} \quad f_L(s) &= M(s) \exp(\sqrt{s}) + N(s) \exp(-\sqrt{s})~.
	\end{align}
\end{subequations}
Finally, Laplace transforming the dynamical equations for the reservoir
concentrations yields
\begin{subequations}
	\label{eq:l2}
	\begin{align}
		f_0(s) &= \frac{1}{s} +\frac{\zeta}{\sqrt{s}}\left(  M(s) - N(s)  \right)\\
		\text{and} \quad f_L(s) &= -\frac{\zeta}{\sqrt{s}}\left( M(s) \exp(\sqrt{s}) - N(s) \exp(-\sqrt{s})\right)~.
	\end{align}
\end{subequations}
Combining the above we obtain explicit solutions for 
$M(s)$ and $N(s)$, thus entirely determining the solutions $f_0(s)$, $f_L(s)$ and $f(s)$ to the problem in the Laplace space.
In particular, for the concentration in the reservoir initially devoid of chemical, we obtain
\begin{widetext}
\begin{equation}
	\label{eq:fL}
	f_L(s) = \frac{  2 \zeta \mathrm{e}^{\sqrt{s}}}  {  \sqrt{s} \left[   
		\left(-1 + \mathrm{e}^{2 \sqrt{s}} \right) s
		+ 2 \zeta \left( 1 + \mathrm{e}^{2 \sqrt{s}}   \right) \sqrt{s}
		+ \zeta^2 \left( -1 +  \mathrm{e}^{2\sqrt{s}}   \right)
		\right]   }~.
\end{equation}
\end{widetext}

This solution in Laplace space is not easily inverted into an analytical expression for the evolution in time of
$c_L(t) = \Lapl^{-1}(f_L)|(t)$. In order to access its time evolution, we adapted a numerical inverse Laplace code in Python \citep{Barbuto} which implements the Zakian method ~\citep{Halsted1972,Abate2006}. The numerical evaluation of 
$c_L(t)$, as a function of the characterisic nondimensional parameter $\zeta = L/\ell$, is shown in figure \ref{fig:res_diff}. It reveals the typical 
nondimensional time-scale of equilibration $1/2 \zeta$, which in dimensional form becomes the previously discussed equilibration time 
$\tau_{\text{eq}} = L \ell / 2 D$. At steady state, the concentration equilibrates between the two reservoirs and the 
tube at a final uniform value {$c_{\text{f}} = 1/ (2+\zeta)$}. 
Finally, for $\zeta \ll 1$, the validity of the approximations of the concentration $c_L(t)$ as a saturating exponential  in equation \eqref{eq:exp_cl} is clearly demonstrated (Figure~\ref{fig:res_diff}, right panel).

\subsection{Mathematical model of remote mutualistic cross-feeding and numerical methods}

\subsubsection*{Membrane limit}

The first natural limit of the model is that of zero channel length $\lambda \to 0$, in which the reservoirs are in contact, but separated by a porous membrane. We call this the {\it membrane limit} because the membrane setup is as in membrane 
experiments \citep{Paul2013}, and we consider instantaneous equilibration of concentrations across the membrane as a good approximation. Fixed 
points for this limit are obtained trivially by letting $\lambda \to 0$ in \eqref{eq:FP2}-\eqref{eq:FP3}, which confirms that metabolite concentrations 
are equalised between reservoirs at steady state. We note that the membrane limit is identical to a \textit{mixed co-culture}, where A and B grow mixed 
together in the same reservoir, except for the dilution effect associated
with the segregation of the two species on either side of the membrane.  The corresponding dynamical system for a mixed 
co-culture also admits a positive fixed point $(a^*,b^*,c^*,v^*)$ under the same conditions \eqref{eq:FP_cond}, with $a^*$ and $b^*$ given by 
\eqref{eq:FP1}, $c^* = c_b^*$ from equation \eqref{eq:FP2} and $v^* = v_a^*$ from equation \eqref{eq:FP3}. As mentioned earlier, such a co-culture model is 
fundamentally different from models considering mutualistic nutrient exchanges implicitly \citep{Murray1989,Yukalov2012,Grant2014a,Holland2010}.  

\subsubsection*{Remotely-fed monoculture \label{sec:remote_mono}}

Another interesting limit is one in which a species in one of the reservoirs is replaced by a fixed concentration of metabolite. For example, we 
could have species B growing on C diffusing through the channel from the remote reservoir. In this limit, the model on the side of C reduces to 
passive diffusion from a source, which provides a useful control on the mutualistic dynamics, as mentioned in the results section.  
The mathematical model for such a remotely-fed monoculture is directly obtained from the remotely cross-feeding populations model (equations~\ref{eq:LD_phy}) by setting one microbial species and the metabolites
it produces to zero.

\subsubsection*{Numerical methods}

The system of non-dimensional equations \eqref{eq:LD_no_dim} is solved numerically through a custom finite difference solver using Python and Cython, based on an explicit centered scheme for the diffusion PDEs and an improved Euler scheme for the integration of the ODEs. The map in Figure \ref{fig:surv_map2} was drawn by setting a minimum threshold concentration of cells below which the mutualistic coculture is considered crashed, here set at \SI{1}{\cells\per\milli\liter} for both species.The coexistence area corresponds to initial concentrations that give rise to a time evolution towards the positive fixed point with cell numbers keeping above the minimum threshold at any time.

\begin{figure*}[tbh!]
	\includegraphics[width=0.9\linewidth]{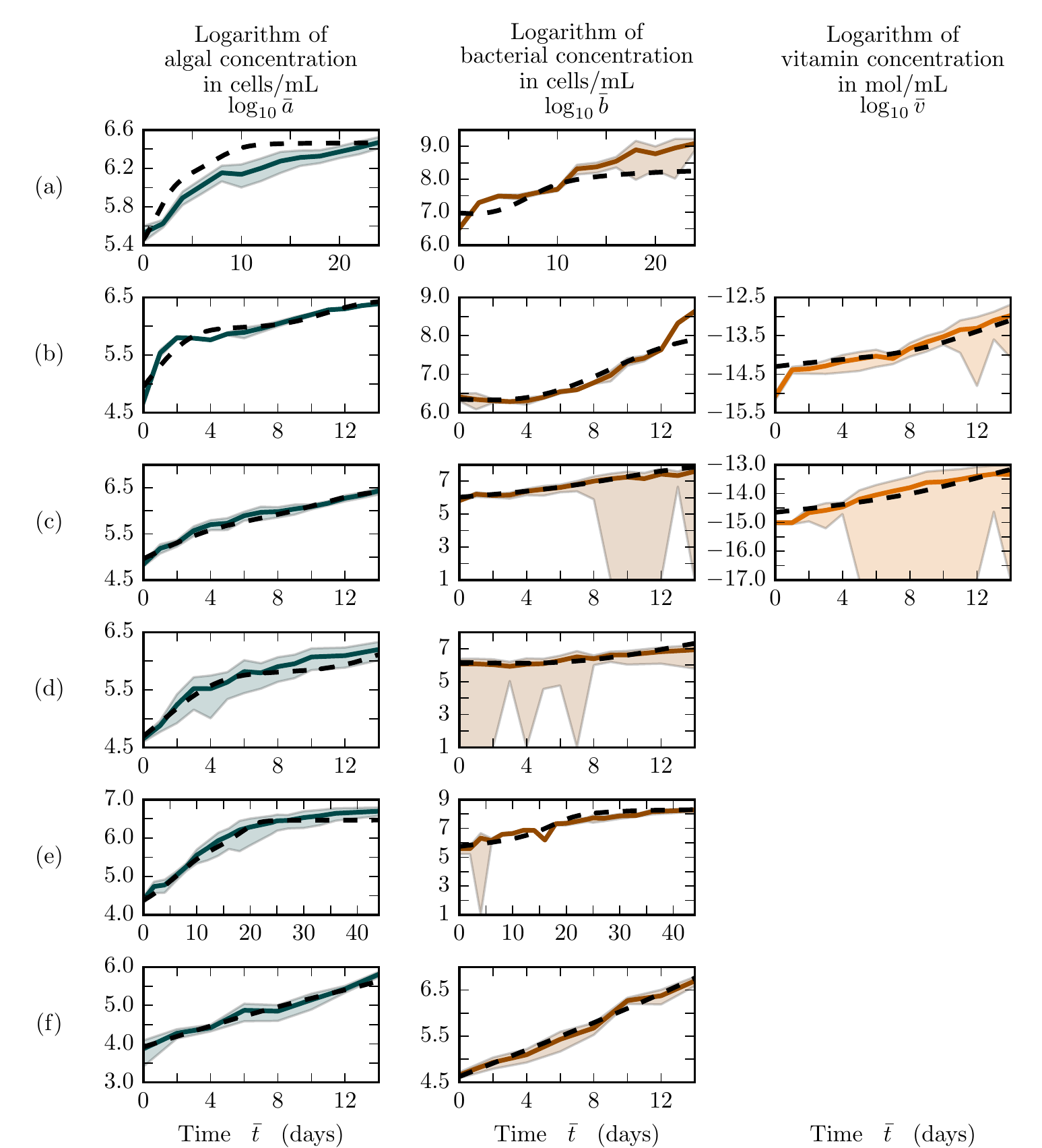}
	\caption{Experimental results and theoretical fits on growth of 
    co-cultures. Rows (a)-(f) display results from 
    6 independent growth experiments for \textit{M.~loti} and \textit{L.~rostrata} cocultures, for different starting values and ratios of the two species. For each experiment, from left to right
    the panels show the algal concentration $\bar{a}$, the bacterial concentration $\bar{b}$, and, when data is available, the vitamin concentration $\bar{v}$. Continuous thick lines show the average value over a set of replicates, with the interval of +/- one standard deviation shown as a shaded area. The fits from the model with parameters from table~\ref{tab:fit_phy_param} are shown with dashed black lines. Number of replicates per experiment from a to f is n = 6, 3, 5, 5, 4 and 4. Large downward shaded areas represent on this logarithmic scale time points for which standard deviation is comparable to the mean.}
	\label{fig:fit_Vai}
\end{figure*}

\subsection{Parameterisation for specific microbial associations}

The results presented in this paper were obtained from numerical studies of the mathematical model with parameter values 
corresponding to the mutualistic association between \textit{Lobomonas~rostrata}, a B$_{12}$-requiring green alga, and 
\textit{Mesorhizobium~loti}, a B$_{12}$-producing soil bacterium \citep{Kazamia2012a}. The following procedure was used to 
obtain these parameter values. First, physiologically relevant ranges for each parameter were collected by direct 
measurement (see next section) or from the published literature. Then, specific parameters -- both nondimensional 
parameters of the reduced model and dimensional parameters to convert experimental data to nondimensional units-- were obtained by minimizing 
the squared distance between simulated time evolution, obtained through a custom finite difference solver in Python, and experimental results on 
mixed cultures, while searching within domains of parameter values 
which contain the physically relevant ones, and validating the fixed-point conditions in equation 
\eqref{eq:FP_cond}. The basin-hopping minimisation procedure gives local optima which capture well the observed dynamics of mixed co-cultures of 
\textit{L.~rostrata} and \textit{M.~loti} (see Figure~\ref{fig:fit_Vai}). The range of physiologically relevant parameters used to constrain the search of parameters for
the association of \textit{M.~loti} and \textit{L.~rostrata} are presented in table \ref{tab:LD_phy_param}, while the fitted parameters,
both dimensional and nondimensional, are given in tables \ref{tab:fit_phy_param2} and \ref{tab:fit_phy_param}. 

\begin{table*}[tbh!]
	\caption{\label{tab:LD_phy_param} Physiologically relevant parameter ranges for the mutualistic association of \textit{M.~loti} and 
	\textit{L.~rostrata}.}
	\sisetup{
		table-number-alignment = center,
		table-figures-integer  = 2,
		table-space-text-post = ~,
		range-phrase = --
	}
	\smallskip 
		\begin{tabular*}{\textwidth}{l @{\extracolsep{\fill}} c r l s[table-unit-alignment = left, table-text-alignment = center] c	}
			\toprule \vspace{-6pt}\\
			{Parameter}  &  {symbol}  &  \multicolumn{2}{c}{value}  & {unit}  &  {source}  \\ 
			\hline \vspace{-6pt}\\
			{Death rate of \textit{M.~loti}} &  $\delta_b$ &  \num{5}  &    $\times~10^{-2}$  & \per\hour  &  \footnote{this work (see SI Estimation of biological parameters)}  \\
			{Diffusivity of carbon (25$^{\circ}$) \footnote{considering glycerol or small sugars such as glucose and sucrose.}} &  $D_c$  & \numrange{1.8}{3.6}  &  $\times~10^{-2}$ & \cm\tothe{2}\per\hour & \cite{Amsden1998}  \\
			{Diffusivity of vitamin B$_{12}$ (\SI{25}{\degreeCelsius})}  &  $D_v$   & \num{1.0} & $\times~10^{-2}$  & \cm\tothe{2}\per\hour & \cite{Amsden1998}  \\
			{Carrying capacity of \textit{L.~rostrata}}  &  $K_a$   &     \numrange{1}{10}  &  $\times~10^{6}$  & \cells\per\ml & \footnotemark[1]  \\
			{Carrying capacity of \textit{M.~loti}}  &  $K_b$  &  \numrange{5}{50}  &    $\times~10^{8}$  & \cells\per\ml & \footnotemark[1]  \\
			{Growth affinity constant of bacteria\footnote{obtained considering \textit{E.~coli} and species of rhizobia growing on different sugars.  
			The range of values is quite wide due to the ability of bacteria to tune their affinity constant depending on the environmental conditions \citep{Ferenci1999}.}} & $K_c$ & \numrange{1}{30000} &  $\times~10^{-10}$  & \mole\per\cm\tothe{3} & 
			\cite{Button1985,Senn1994}  \\
			{Growth affinity constant of algae\footnote{obtained considering \textit{L.~rostrata} and other B$_{12}$-dependent species.}} & $K_v$ & \numrange{1}{100} &  $\times~10^{-16}$ & \mole\per\cm\tothe{3} & \cite{Button1985, Droop2007}  \\
			{Maximum growth rate of \textit{L.~rostrata}} &  $\mu_a$ & \num{1.25} &    $\times~10^{-2}$   & \per\hour  & \cite{Kazamia2012a}  \\
			{Maximum growth rate of \textit{M.~loti}} &  $\mu_b$ & \numrange{1}{2} & $\times~10^{-1}$  & \per\hour  & \footnotemark[1] \\
			{Release rate of carbon by algae\footnote{obtained considering two species belonging to the same family (\textit{Chlamydomonadaceae}) as \textit{L.~rostrata}, and arabinose molar mass.}}  &  $p_c$  &  \numrange{1}{100}  &  $\times~10^{-16}$  & \mole\per\cells\per\hour  
         & \cite{Miller1974,Boyle2009,Kamjunke2009} \\
			{Release rate of vitamin by bacteria\footnote{obtained considering two B$_{12}$-producing bacterial species, \textit{Azobacter vinelandii} and \textit{Halomonas} sp. }}   &  $p_v$  &  \numrange{1}{50}  &  $\times~10^{-23}$   & \mole\per\cells\per\hour  
         & \cite{Gonzalez-Lopez1983,Croft2005} \\
			{Yield of algae over \BXII{}}  &  $Y_a$  &  \numrange{1}{100}  &    $\times~10^{20}$   & \cells\per\mole  & $K_a/K_v$ \\
			{Yield of bacteria over organic carbon}  &  $Y_b$  &  \numrange{1}{e6} &  $\times~10^{13}$  & \cells\per\mole  &  $K_b/K_c$, \cite{Link2008} \\
			\botrule
		\end{tabular*}
\end{table*}

\begin{table*}[tbh!]
	\caption{Fitted parameters for the mutualistic association of \textit{M.~loti} and \textit{L.~rostrata}.}
	\centering
	\label{tab:fit_phy_param2}
	\setlength{\tabcolsep}{5pt}
	\smallskip
	\begin{tabular}{l c c}
		\toprule \vspace{-6pt}\\
		Fitted dimensional parameter              &         Symbol          &   Value \\
		\hline \vspace{-6pt}\\
		Algal carrying capacity                       &    \( K_a \)    &      \SI{3.0e6}{\cells\per\milli\liter}       \\
		Bacterial carrying capacity                  &     \(K_b\)     &      \SI{5.8e8}{\cells\per\milli\liter}      \\
		Growth affinity constant of algae       &    \(K_v\)         &    \SI{1.2e-14}{\mole\per\centi\meter\tothe{3}}       \\
		Maximum growth rate of \textit{M.~loti}        &         \(\mu_b\)           &     \SI{1.9}{\per\hour}      \\
		\botrule
	\end{tabular}
\end{table*}

\subsection{Estimation of biological parameters}

\subsubsection*{\it Monoculture experiments: Carrying capacities of \textit{M.~loti} and \textit{L.~rostrata}}

Liquid cultures of \textit{M.~loti} were grown for 3 days ( \SI{33}{\degreeCelsius}, shaken at 240 rpm) in TY 
medium (tryptone \SI{5}{\gram\per\liter}, yeast extract \SI{3}{\gram\per\liter}, \ce{CaCl2*2H2O} \SI{0.875}{\gram\per\liter}) and washed in TP+ before serial dilution for counting of colony forming units.
The post-wash concentration was estimated to be $5-$\SI{10e8}{\cells\per\milli\liter}. Given the existing loss of cells during washing, we therefore allow the bacterial carrying capacity of our model $K_b$ to be in the range $5-$\SI{50e8}{\cells\per\milli\liter}.
Similarly, we estimated the carrying capacity of \textit{L.~rostrata} by growing these algae in TP+ with \SI{100}{\nano\gram\per\liter} of vitamin B$_{12}$ for 6 days to saturation (\SI{22}{\degreeCelsius}, shaken at 200 rpm, day/night cycle of 14h/10h), and plating them after washing in TP+ and serial dilution on TY agar plates for colony forming unit counting. We recorded saturation concentration $\sim$\SI{2e6}{\cells\per\milli\liter}, which, allowing for losses during cell washing, results in an accepted range of $1-$\SI{10e6}{\cells\per\milli\liter} for the algal carrying capacity $K_a$ in our model.

\subsubsection*{\it Monoculture experiments: Death rate of \textit{M.~loti}} A pre-culture of \textit{M.~loti} in TY as above was washed in fresh TP+ and inoculated at a concentration $b_0 = \SI{3.2e8}{\cells\per\milli\liter}$ in \SI{70}{\milli\liter} of TP+ without carbon source. Every two days, a \SI{100}{\micro\liter} sample was taken to determine a live cell concentration through counting of colony forming units (CFUs) on TY agar. After a 2 days lag period, we measured an exponential decay of the bacterial population with death rate $\delta_b \approx \SI{5e-2}{\per\hour}$ over the next 6 days.

\subsubsection*{\it Co-culture experiments:  Global fit of model parameters}

The experiments whose outcomes were used to fit the model parameters utilised
the following protocol.
\textit{L.~rostrata} and \textit{M.~loti} were grown in TP+ medium at \SI{25}{\degreeCelsius} on a 12h/12h day/night cycle, with 100 microeinsteins of light and shaking at 120 rpm. Bacterial concentrations were estimated with counts of CFUs on TY 
agar, and algal concentrations were obtained with a Coulter counter. In some experiments, B$_{12}$ concentration was estimated with bioassays \citep{Raux1996}.
Figure \ref{fig:fit_Vai} shows the results for a set of six independent 
experiments (a-f) along with global fits to the model, corresponding
to the values shown in Table~\ref{tab:fit_phy_param}.

\subsection{Mutualism at a distance: experimental proof of concept}

\begin{figure*}[tbh!]
	\centering 
	\includegraphics[width=0.7\linewidth]{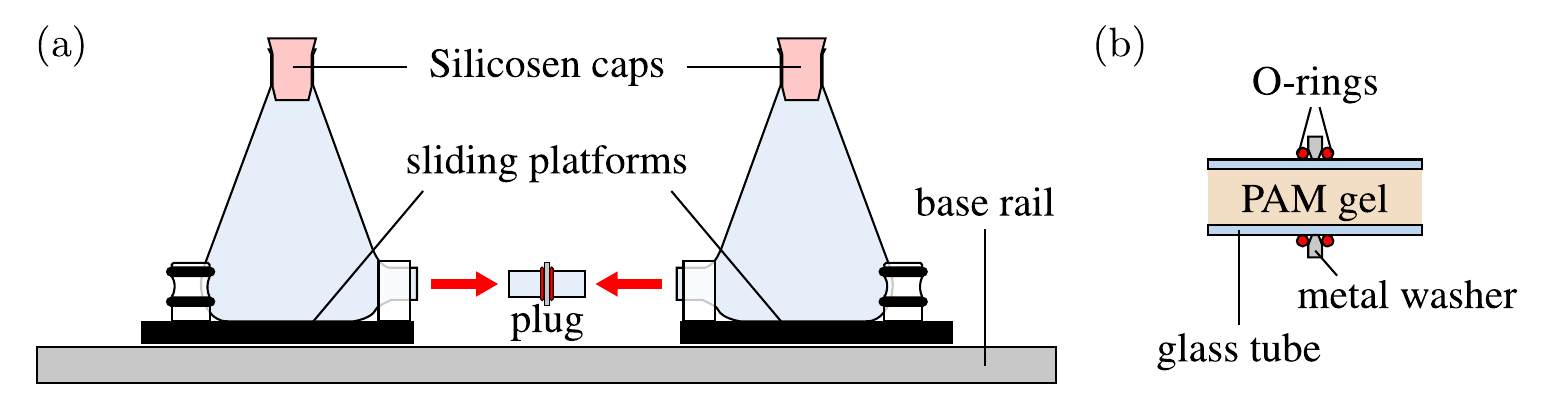}
	\caption{Chambers for proof-of-principle experiments. (a) Sketch of the platform holding the modified flasks during assembly. (b) Sketch of the diffusive plug filled with polyacrylamide (PAM) gel, used to connect the two flasks in experiments of mutualism at a distance. }
	\label{fig:sketch_connectedflasks}
\end{figure*}

To test experimentally the predictions of the mathematical model, we developed a system to culture mutualistic microbial species exchanging metabolites diffusively over a finite distance. Briefly, each of two \SI{100}{mL} conical Erlenmeyer flasks was modified  (Soham Scientific Ltd) to have a side arm (\SI{8}{\milli\meter} long, outside diameter \( \SI{11}{\milli\meter}\), inside diameter \(\SI{9}{\milli\meter}\)) in which a small glass tube could be inserted (\SI{25}{\milli\meter} long, outside diameter \SI{8.65}{\milli\meter}, inside diameter \SI{7.45}{\milli\meter}). Sealing of the tube-flask junction was achieved 
by compression of O-rings on each side of a metal washer glued onto the glass tube (see figure~\ref{fig:sketch_connectedflasks}a,b). The
force of compression was established and maintained by mounting the flasks on custom sliding platforms (figure~\ref{fig:sketch_connectedflasks}b,c). To prevent contamination, flasks were capped with silicon plugs (Hirschmann Silicosen type T-22) and aluminium foil, while the middle area of the flasks and tube assembly was also further covered with aluminium foil. 
The central glass tube connecting the inside of both flasks was filled with a polyacrylamide (PAM) gel (4\% acrylamide w/v  with a relative concentration of bis-acrylamide of 2.7\%, filter-sterilised before pouring, BioRad). Once polymerised, the gels in their tubes were put in a bottle of sterile water and left to soak for 6 days to allow for any of the toxic non-polymerised monomer to diffuse out of the gel. We verified the very weakly hindered diffusion of \BXII{} through this gel by colorimetry, measuring a reduction of $\sim 10\%$ of diffusivity with respect to \BXII{} diffusion in water, which validates the chosen gel pore size as allowing the diffusive transport of small metabolites. We also performed a test to check for cross-migration of the mutualistic species. Both flasks were filled with a rich bacterial medium for soil bacteria (TY), but only one side was inoculated with \textit{M. loti} (see below for strain details). These bacteria reached a saturation density within a few days, but over a timescale of $2.5$ months no bacteria were detected in the first flask, proving the PAM gel is not penetrable by bacteria (and by inference by the algae, which are larger). 

In such connected flasks, we inoculated one side with the \BXII{}-dependent green alga \textit{Lobomonas rostrata} (SAG 45-1, wild type strain) and the other with the \BXII{} producing bacterium \textit{Mesorhizobium loti} (MAFF 303099, wild type strain, original gift from Prof. Allan Downie,  John Innes Centre, UK). Both inocula were diluted with TP+ medium~\citep{Kazamia2012a} to the desired starting concentrations of microbes. The \textit{L. rostrata} pre-culture was grown in TP+ with \SI{100}{\nano\gram\per\liter} of vitamin \BXII{} 
from colonies picked from a slant, while the \textit{M. loti} pre-culture was grown in TY medium. Both pre-cultures were washed in fresh TP+ before inoculation in the assembly in order to remove any organic carbon and \BXII{} in the initial growth media. The initial concentrations of \textit{M. loti} and \textit{L. rostrata} were \(b_0 = \SI{2.2e8}{\cells\per\milli\liter}\)  and \(a_0 = \SI{5.3e4}{\cells\per\milli\liter}\), inferred from viable counts. To ensure culture sterility, flask assembly and inoculation were carried out in a laminar biosafety cabinet (PURAIR VLF 48). 
The connected flasks were mounted on a shaking platform (120rpm) 
within an incubator for 50 days, at \SI{25}{\degreeCelsius}, with continuous illumination (\SI{80}{\micro\mole\per\meter\tothe{2}\per\second}). After this period, these assemblies were left in static incubation at \num[separate-uncertainty = true]{20(2)}~\si{\degreeCelsius} and at ambient day/night light
levels.

\subsubsection*{\it Viable counts and \BXII~concentration measurements \label{sup:prelim_exp_results}}

Algal and bacterial populations were sampled $55$ and $230$ days after inoculation. No contamination (external or between species) was detected, and PCR screening was used to confirm species identity as \textit{ Mesorhizobium loti} bacteria and \textit{Lobomonas rostrata} algae. This confirms the ability of the PAM gel to prevent cells from crossing, while allowing metabolites to be exchanged.

Viable counts revealed that the population of bacteria $55$ days after inoculation was $\sim 10^3$ smaller than the inoculum. At the same time point the algae had grown little: the cell concentration was only $1.3$ times larger than the inoculum. After $230$ days
the bacteria had recovered, and the algae had grown significantly. 
At this time the algal concentration from two replicates was $a=7.8\pm0.3\times 10^5$ cells/cm$^3$ (where the uncertainty is the standard error in the mean), about $15$ times the inoculation concentration and close to the carrying capacity they reach in well-mixed co-cultures (see table~\ref{tab:LD_phy_param}). While slight initial growth of the algae might be attributed to internal reserves of vitamin \BXII, it is difficult to account for growth $230$ days after inoculation in the absence of the vitamin. Indeed, using bioassays \citep{Raux1996} we measured a \BXII{} concentration of $24\pm3$ pg/ml in the medium on the side of the algae. On the side of the bacteria, we found $132\pm7$ pg/ml. This implies the existence of a concentration gradient across the tube between the two flasks. This is required for the supply of the \BXII{} to the algae, as predicted by the model (see equation \ref{eq:FP3}).
 
\begin{figure}[tbh!]
	\centering 
	\includegraphics[width=1.0\linewidth]{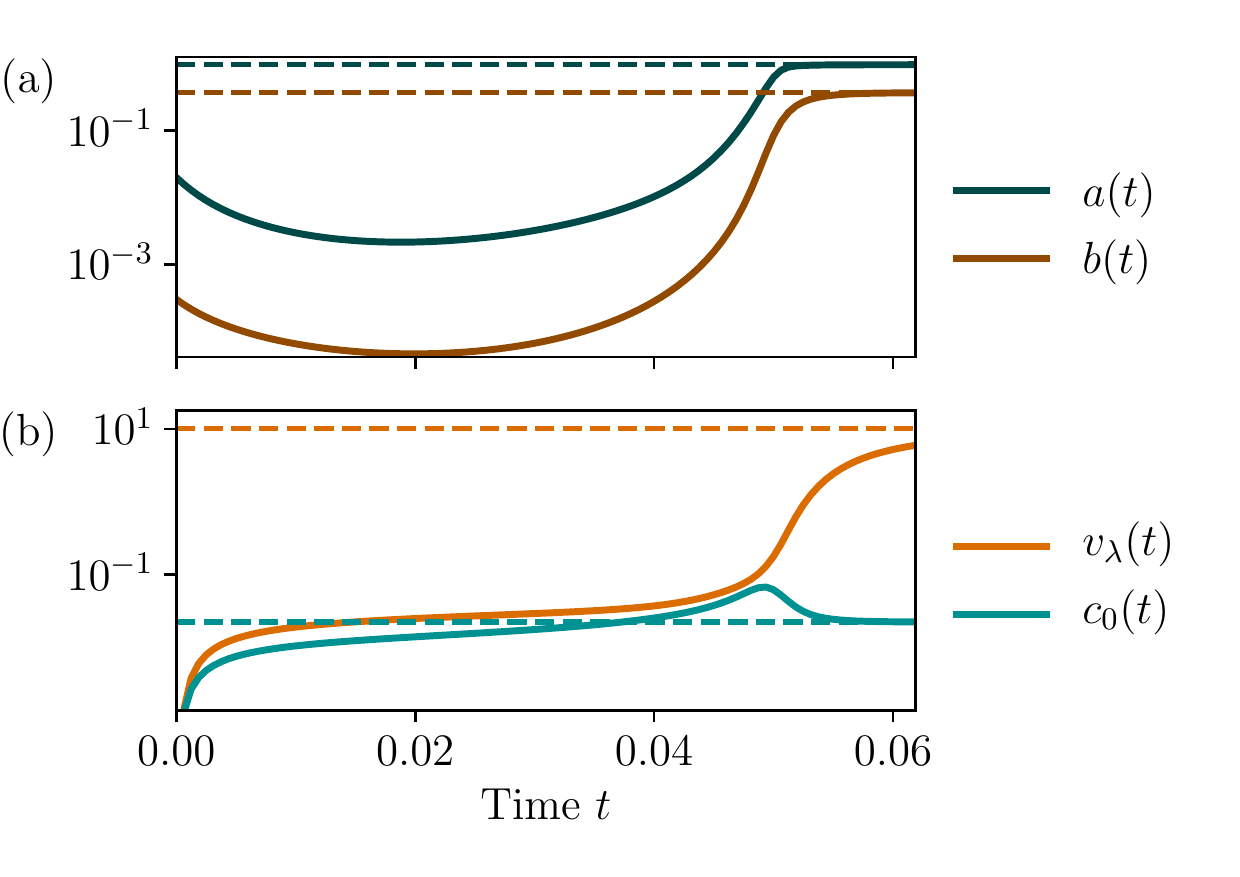}
	\caption{Example of oscillations of (a) concentrations of cells and (b) concentration of metabolites during the time evolution of a co-culture at a distance system before convergence. Initial parameters are close to the boundary between survival and extinction ($\lambda = 2$, $\eta = 3$, $a_0 = \num{2e-2}$, $b_0 = \num{3e-4}$ and no initial nutrients).\label{fig:time2SS_sup}}
\end{figure}

\bibliography{library_new}

\end{document}